\documentclass[aps,prc,preprint,groupedaddress,showpacs]{revtex4-1}
\usepackage{graphicx}
\usepackage{amsfonts}
\usepackage{amsmath}
\usepackage{amssymb}
\usepackage{bm}
\usepackage{afterpage}
\usepackage{natbib}
\newcommand{\pr}{\prime} 
\newcommand{\ket}[1]{\left| #1 \rangle\right.} 
\newcommand{\bra}[1]{\langle #1 |} 
\newcommand{\braket}[2]{\langle #1 \vphantom{#2} |
  #2 \vphantom{#1} \rangle} 
\newcommand{\ketbra}[2]{| #1 \vphantom{#2} \rangle 
  \!\ \!\langle #2 \vphantom{#1} |} 
\newcommand{\matrixel}[3]{\langle #1 \vphantom{#2#3} |
 #2 | #3 \vphantom{#1#2} \rangle} 

\begin{document}
\title{Real-Space Imaginary-Time Propagators for
Non-Local Nucleon-Nucleon Potentials}
\author{J.E. Lynn}
\email{joel.lynn@asu.edu}
\author{K.E. Schmidt}
\email{kevin.schmidt@asu.edu}
\affiliation{Department of Physics,
Arizona State University, Tempe, Arizona, 85287, USA}
\date{\today}
\begin{abstract}
Nuclear structure quantum Monte Carlo methods such as
Green's function or auxiliary field diffusion Monte Carlo
have used 
phenomenological local real-space potentials containing as few derivatives 
as possible, such as the Argonne-Urbana family of interactions,
to make 
sampling simple and efficient.  Basis set methods such as no-core shell 
model and coupled-cluster techniques typically use softer non-local 
potentials because of their more rapid convergence with basis set size. 
These non-local potentials are usually defined in momentum space and are 
often based on effective field theory. Comparisons of the results of the two
types of methods can be difficult when different potentials are used.  We 
show methods for evaluating the real-space imaginary-time propagators 
needed to perform quantum Monte Carlo calculations using such non-local 
potentials. We explore the universality of the large imaginary time 
propagators for different potentials and discuss how non-local potentials
can be used in quantum Monte Carlo calculations.
\end{abstract}

\maketitle
\section{Introduction}
Quantum chromodynamics (QCD) is the fundamental theory that underlies the
description of atomic nuclei, where quarks and gluons are the
degrees of freedom.  However, despite four decades of studying QCD, 
little connection has been made between QCD and low-energy many-body 
nuclear dynamics.
The most direct computational application of 
QCD, lattice QCD, has made much progress in the past decades but remains 
some distance away from being a practical tool for computing many-body 
nuclear observables.  For recent reviews of the outlook for lattice QCD 
calculations as they apply to nuclear physics see Refs. 
\cite{savage2010,beane2011}.
While direct applications of effective field theory have made
progress \cite{lee2009,epelbaum2010,epelbaum2011,epelbaum2012} 
the characterization of atomic nuclei in terms of phenomenological
two- and three-body
nucleon interactions remains the standard starting
point for most nuclear structure
calculations.

Quantum Monte Carlo is one of the
most successful methods for nuclear matter and nuclear structure
calculations. Green's function Monte Carlo has solved for many
low lying states of nuclei for
$A \le 12$ \cite{carlson1987,Pieper:2001mp}
and auxiliary-field diffusion Monte
Carlo can calculate much larger nuclei and nuclear and neutron 
matter \cite{schmidt1999,gandolfi2007,gandolfi2007b}.  These 
methods have used phenomenological local real-space
potentials containing as few derivatives as possible, such as the 
Argonne-Urbana family of 
interactions \cite{Wiringa:1994wb,Carlson:1983kq,Pieper:2001ap}, to make 
sampling easy and efficient.
See Ref. \cite{Pieper:2001mp} for a review of Green's function Monte 
Carlo results.

However, there are other successful approaches that can reach $A=12$ and 
beyond.  Basis-set methods such as the no-core shell 
model \cite{navratil2000,navratil2009} and 
coupled-cluster techniques \cite{hagen2010,jansen2011,hagen2008} have 
typically used softer non-local potentials as these have more rapid 
convergence with basis-set size.  These potentials are typically defined in
momentum space and are often derived from chiral effective field theory 
such as the next-to-next-to-next-to-leading order (N$^3$LO) interaction of 
Ref. \cite{Entem:2003ft}.

Comparison of the results of basis set methods with Monte Carlo
calculations can be difficult when different 
potentials are used. Quantum Monte Carlo methods use propagation in
imaginary time to project out the low-energy states of a quantum
many-particle system. The propagation is performed by first writing the
many-body short-imaginary-time propagator. Accurate methods
use a pair-product approximation \cite{ceperley1995,pudliner1997} where the
many-body propagator is written in terms of the propagator for each pair of
particles. In order to perform a quantum Monte Carlo calculation using
an arbitrary pair potential, we need to calculate the pair propagator in
imaginary time.

The aim of this paper is to show how to
evaluate this real-space imaginary-time pair propagator needed to perform 
quantum Monte Carlo calculations using non-local potentials 
(Sec. II), to demonstrate the consistency of the propagators using 
different potentials at large imaginary times (Sec. III), and to discuss
how to use these propagators to calculate the properties of
nuclei and nuclear matter using these non-local potentials
with quantum Monte Carlo methods
(Sec. IV).   

\section{Methods}
Non-local potentials are typically generated in momentum space from an
effective field theory.  With a potential defined in momentum space,
$V(k,k^{\pr})$, it is natural that we proceed by constructing a Hamiltonian
in momentum space as well. Our normalization and completeness conventions 
for our continuous real-space and momentum-space basis states
are
\begin{equation}
1=\int d^3r\ketbra{\bm{r}}{\bm{r}} =
\int\frac{d^3k}{(2\pi)^3}\ketbra{\bm{k}}{\bm{k}}.
\end{equation}
$\bm{r}$ is the separation vector of the two nucleons and 
$\bm{p}=\hbar\bm{k}$ the conjugate momentum.  The overlaps between the 
states are
\begin{equation}
\braket{\bm{r}}{\bm{k}}=e^{i\bm{k}\cdot\bm{r}}.
\end{equation}
We
work in the standard channel basis where $J^2$, $J_z$, 
$L^2$, $S^2$, $S_z$, $T^2$ and $T_z$ are good quantum numbers, with
the total spin,
$\bm{S}=\bm{S}_1+\bm{S}_2$, the total angular momentum
$\bm{J}=\bm{L}+\bm{S}$, and total isospin
$\bm{T}=\bm{T}_1+\bm{T}_2$. 
We choose our basis states as $\ket{rJMLSTT_z}$ and $\ket{kJMLSTT_z}$, with
normalization and completeness given by (suppressing $J$, $S$, $T$, and 
$T_z$)
\begin{eqnarray}
1&=&\sum_{LM}\int_{0}^{\infty}r^2dr\ketbra{rLM}{rLM}
\nonumber\\
&=&\sum_{LM}\int_{0}^{\infty}\frac{k^2dk}{(2\pi)^3}\ketbra{kLM}{kLM}.
\end{eqnarray}
The overlaps are 
\begin{equation}
\braket{rLM}{kL^\pr M^\pr}=4\pi i^Lj_L(kr)\delta_{LL^\pr}\delta_{MM^\pr}.
\end{equation}

For numerical work, we compactify our real space to a sphere 
of radius $R$.  We choose the Dirichlet boundary condition on the sphere
which forces our momentum-space spectrum to be discrete with 
$k_n^{(L)}R$
being the zeros of the spherical Bessel functions, $j_L(k_n^{(L)}R)=0$. 
Below we often drop the superscript $(L)$ when its value is clear from
context.
The discrete momentum states $\ket{k_nLM}$ are chosen with unit
normalization so that 
\begin{equation}
\braket{rLM}{k_nL^\pr M^\pr}=\sqrt{\frac{2}{R^3j_L^\pr(k_nR)^2}}
j_L(k_nr)
\delta_{LL^\pr}\delta_{MM^\pr}.
\end{equation}
Our transformations can now be treated as orthogonal-matrix multiplications.

The momentum-space Hamiltonian for the uncoupled channels where $L=J$  (for
a given set of the quantum numbers --- $J$, $M$, $L$, $S$, $T$, and $T_z$ --- 
which we suppress below) is
\begin{equation}
\matrixel{k_m}{H}{k_n}=\frac{\hbar^2 k_n^2}{2m_r}\delta_{mn}+V(k_m,k_n),
\end{equation}
with $m_r$ the reduced mass.  For the coupled channels where the potentials
couple the $L=J\pm1$ states together the Hamiltonian is
\begin{widetext}
\begin{equation}
\langle k_mL|H|k_nL^\prime\rangle=\left(\begin{array}{cc}
\frac{\hbar^2k_n^{(-)\,2}}{2m_r}\delta_{mn}+V_{--}(k_m^{(-)},k_n^{(-)})
&V_{-+}(k_m^{(-)},k_n^{(+)})\\
V_{+-}(k_m^{(+)},k_n^{(-)})&
\frac{\hbar^2k_n^{(+)\,2}}{2m_r}\delta_{mn}+V_{++}(k_m^{(+)},k_n^{(+)})
\end{array}\right),
\end{equation}
\end{widetext}
where the superscripts $(-)$ and $(+)$ correspond to $L$ or $L^\prime$
having values of $J-1$ and $J+1$.
We then construct the 
momentum-space, imaginary-time propagator by diagonalization of the 
Hamiltonian, giving
\begin{equation}
\matrixel{k_m}{e^{-H\tau}}{k_n}=\sum_{i=1}^{N_k}\braket{k_m}{\psi_i}
e^{-E_i\tau}\braket{\psi_i}{k_n},
\end{equation}
for the uncoupled channels, and
\begin{equation}
\matrixel{k_mL}{e^{-H\tau}}{k_nL^\prime}=
\sum_{i=1}^{N_k}\braket{k_mL}{\psi_i}e^{-E_i\tau}
\braket{\psi_i}{k_nL^\prime},
\end{equation}
for the coupled channels.
The $\{\ket{\psi_i}\}$ are eigenvectors of the Hamiltonian with 
corresponding eigenvalues $\{E_i\}$.  $N_k$ is the number of discrete 
momentum states we keep.  We ensure that $N_k$ is large enough such that 
the propagators converge.  An estimate of how large 
$k_{\text{max}}$ should be can be given by considering the kinetic energy 
alone.  We want $k_{\text{max}}$ such that 
$\exp{\left(-\frac{\hbar^2k_{\text{max}}^2}{2m}\tau\right)}$ can be 
neglected.  In practice, we check the convergence by doubling 
our estimate for $k_{\text{max}}$ and ensuring our results do not change to
the desired precision.  With these methods, it is easy to ensure that the
numerical truncation errors are completely negligible.  For example, for the
results shown here, we use $k_{\text{max}}=40\text{ fm}^{-1}$ 
($N_k\sim80$), and the propagators have truncation errors less than 
$10^{-10}$.

After transforming to real space, we have the matrix elements
\begin{equation}
\matrixel{rJMLSTT_z}{e^{-H\tau}}{r^{\pr}JML^{\pr}STT_z}.
\end{equation}
However, for use in Monte Carlo codes, we want the propagators in a 3D 
real-space basis, $\ket{r\theta\phi SM_STT_z}$,  
\begin{equation}
\ket{r\theta\phi SM_STT_z}=\sum_{JMLM_L}
C^{JM}_{SM_SLM_L}Y_{LM_L}(\theta,\phi)\ket{rJMLSTT_z}.
\end{equation}
$C$ is a Clebsch-Gordan coefficient, $Y$ a spherical harmonic.  

In quantum Monte Carlo calculations, the particle positions are typically
sampled from the central part of the propagator.  The non-central parts are
then included in the spin-isospin samples (auxiliary field diffusion 
Monte Carlo) or sums (Green's function Monte Carlo).  We will 
sample the propagators for these non-local potentials in the same way.  
Here, we define the central part of the propagator as the trace over all 
spins and isospins.
For convenience we also choose a particular coordinate system where the 
initial separation lies along the $z$ axis, and
the final separation is in the $xz$ plane such that we may take 
$\theta=\phi=\phi^{\pr}=0$ and we can visualize the central part of the
propagator as a function of $r-r^{\pr}$ and $\theta^{\pr}$.  For any 
particular application, we can always rotate into this configuration, 
propagate, and rotate back.  The central part of the propagator is then 
written as
\begin{equation}
\label{eq:centralprop}
G(r,r^{\pr},\theta^{\pr};\tau)=\sum_{SM_STT_z}
\matrixel{rSM_STT_z}{e^{-H\tau}}{r^{\pr}\theta^{\pr}SM_STT_Z}.
\end{equation}

\section{Consistency of the propagators at large imaginary times}
In the limit of large imaginary times, we expect that the propagators for 
different potentials should agree.  The propagators are essentially density
matrices for the two nucleon system:
\begin{equation}
\label{eq.densitym}
\rho=\frac{\sum_i\ket{\psi_i}e^{-E_i\tau}\bra{\psi_i}}
{\sum_ie^{-E_i\tau}};\qquad\text{tr}\rho=1,
\end{equation}
corresponding to thermal equilibrium at the temperature $k_BT=\tau^{-1}$.
Now, since any measurable quantity can be written as an
expectation of a Hermitian operator $O$ which can be obtained via 
$\langle O\rangle=\text{tr}(\rho O)$, the density matrices (propagators) we
obtain for the various potentials contain all the
measurable
information for this system. If the position or momentum of the nucleons 
could be
determined with arbitrary precision, the density matrix would be in 
principle measurable. Since the position and momentum are not well defined 
for arbitrary values, the propagator is not completely measurable.
If the various potentials we use are phase-shift 
equivalent --- meaning they reproduce the physical scattering data at or below
$E_{\text{lab}}\approx350\text{ MeV}$ ($E_{\text{c.m.}}\approx175$ MeV) --- then
we would expect that starting at imaginary times 
$\tau\approx(175 \text{ MeV})^{-1}$, the various propagators should begin to
agree more and more.  In fact, we find that for 
$\tau\approx(50\text{ MeV})^{-1}$, the 
higher-energy modes not constrained by current experimental data do not
contribute substantially to the propagator.

We see from Eq. (\ref{eq.densitym}) that at
$\tau\approx(175\text{ MeV})^{-1}$, energies of 175 MeV and above are
suppressed by a factor of $1/e$.  Therefore, 
it is not surprising that at 
$\tau\approx(50\text{ MeV})^{-1}$, where energies of 175 MeV and above are
suppressed by a factor of $1/e^3\approx0.0498$ we find relatively good
agreement between the propagators with different potentials.  This result is
analogous to the renormalization group results leading to the  
$V_{\text{low }k}$ potential of Ref. \cite{bogner:2001gq} and the 
similarity 
renormalization group results of Ref. \cite{bogner:2006pc},
where the high energy modes are integrated out.  
	
\begin{figure*}[htp]
	\vspace{1cm}
	\includegraphics{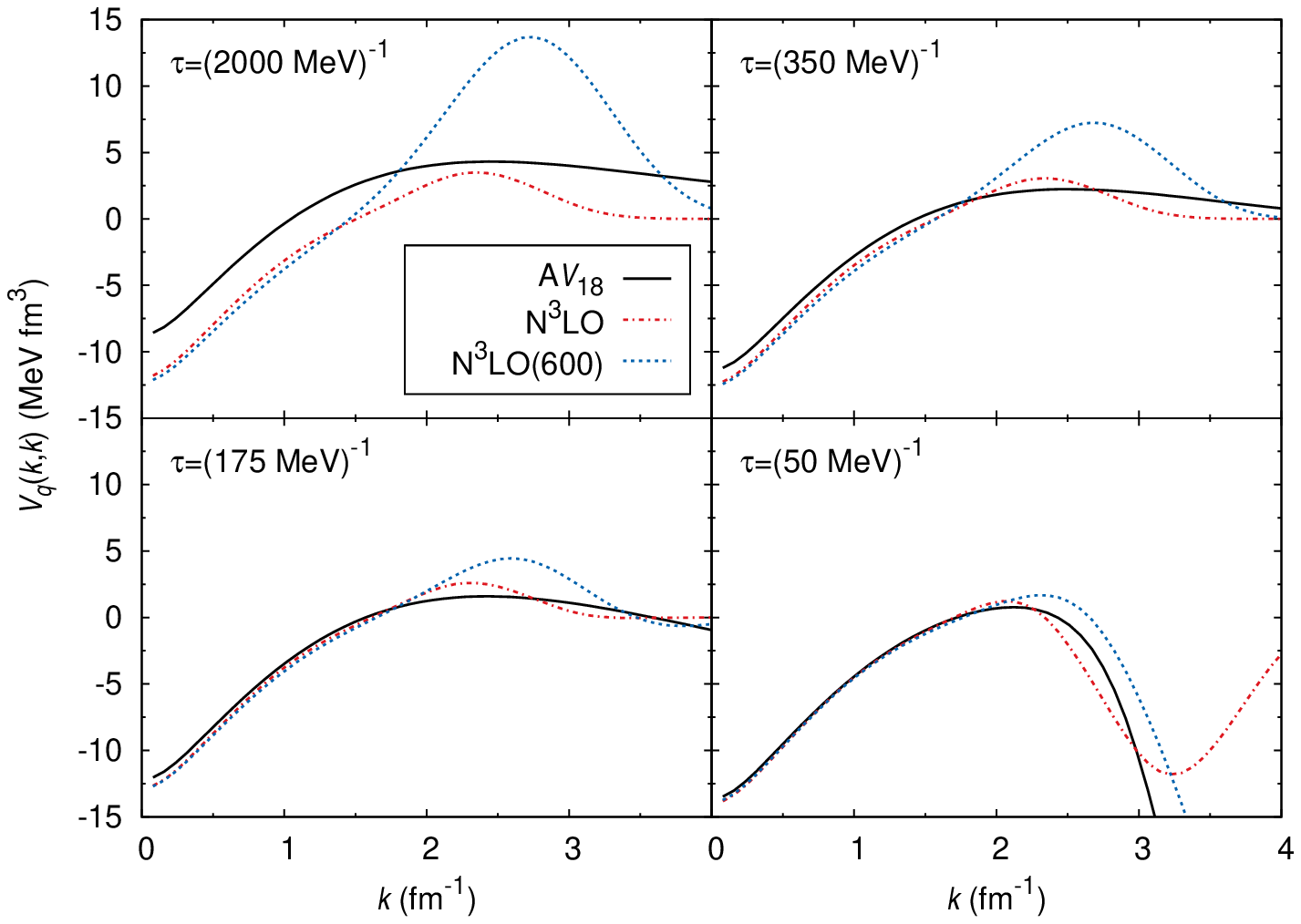}
	\vspace{1cm}
	\caption{\label{fig:1S0diag}(color online) The quantum potential in the
	$^1\!S_0$ partial wave in momentum space as the propagator is calculated
	for successively longer imaginary times.}
\end{figure*}

\begin{figure*}[htp]
	\vspace{1cm}
	\includegraphics{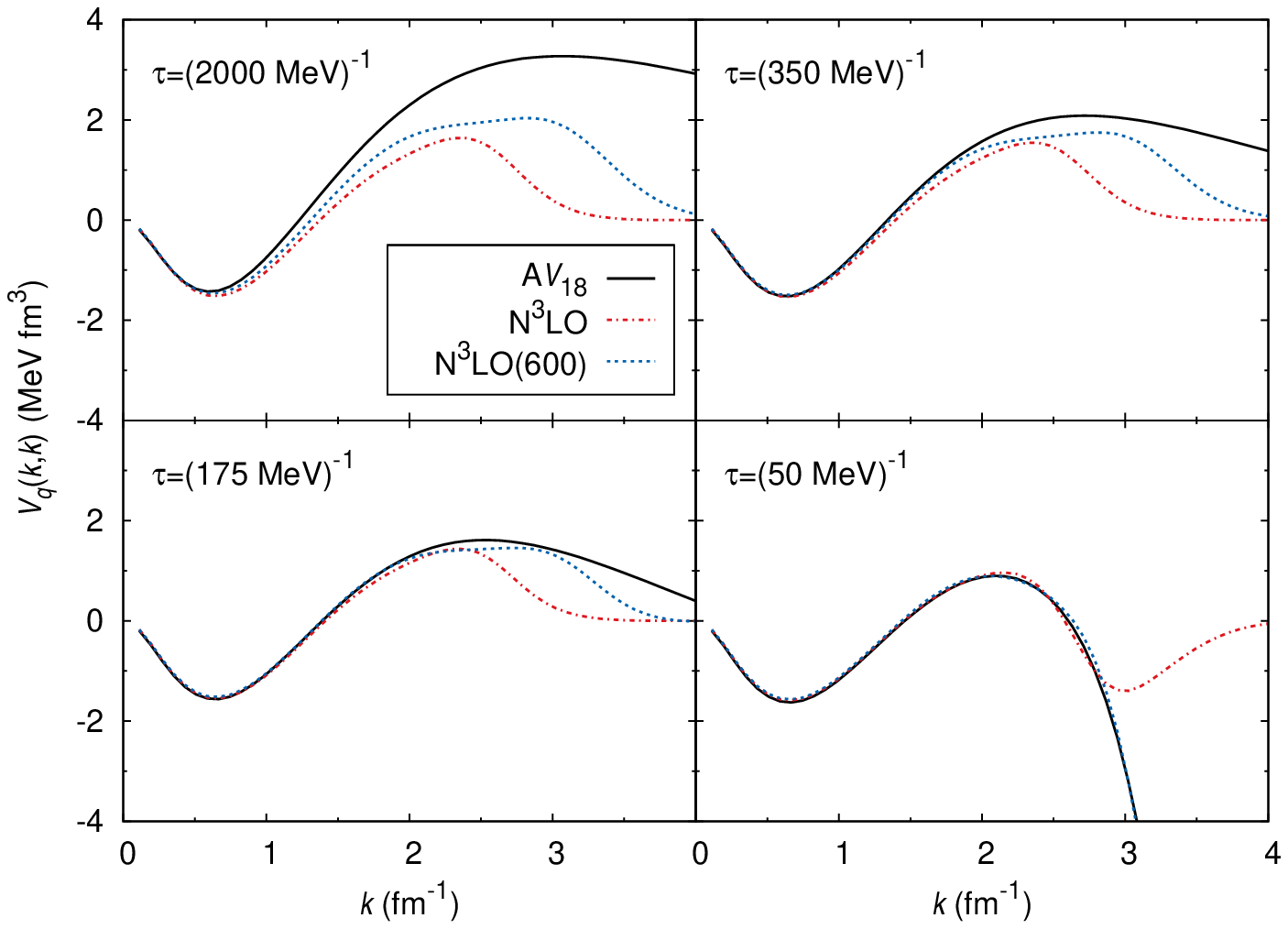}
	\vspace{1cm}
	\caption{\label{fig:3P0diag}(color online) The quantum potential in the 
	$^3\!P_0$ partial wave in momentum space as the propagator is calculated
	for successively longer imaginary times.}
\end{figure*}

\begin{figure*}[htp]
	\vspace{1cm}
	\includegraphics{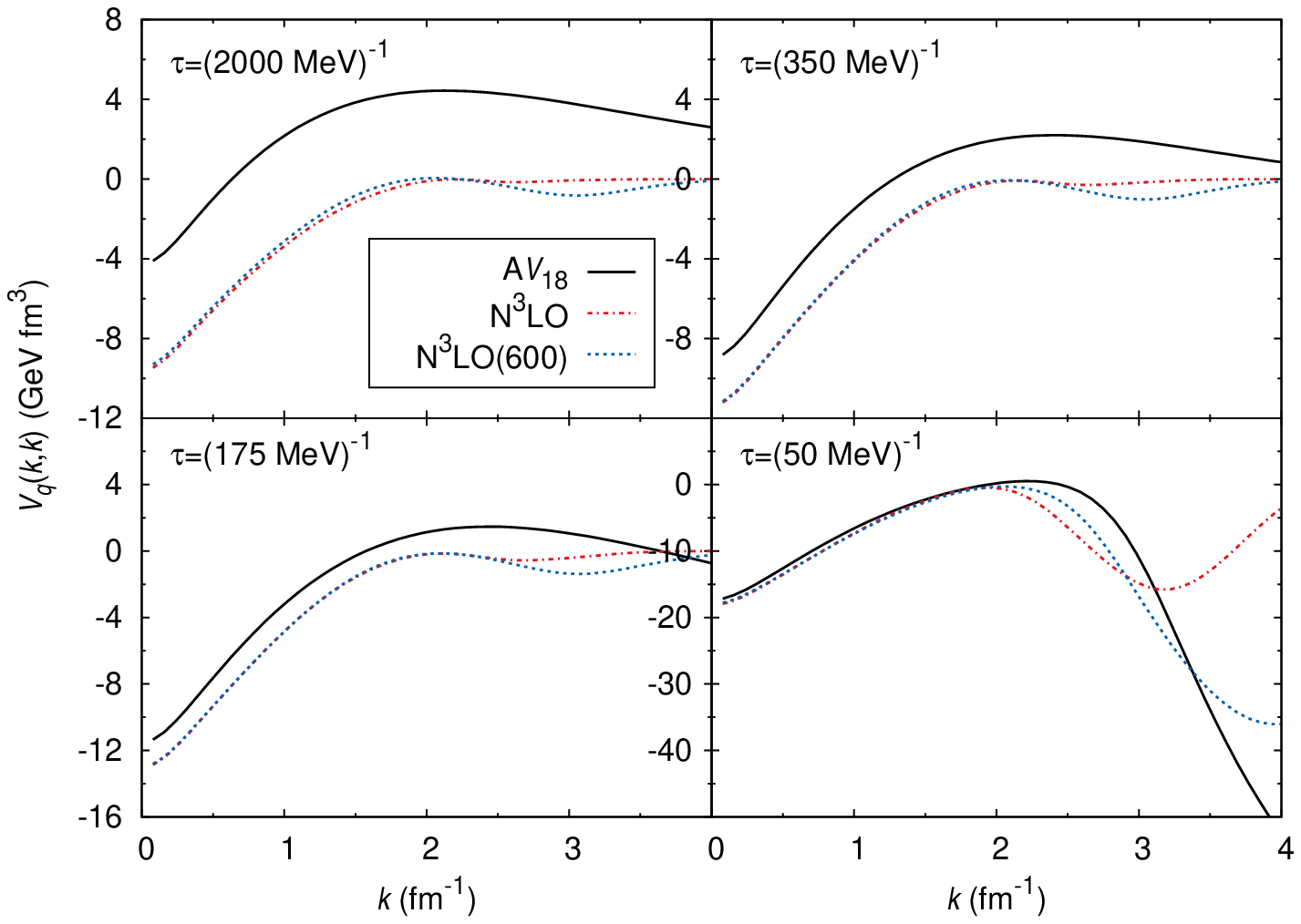}
	\vspace{1cm}
	\caption{\label{fig:3S1diag}(color online) The quantum potential in the 
	$^3\!S_1$ partial wave as the propagator is calculated for successively 
	longer imaginary times.}
\end{figure*}

\begin{figure*}[htp]
	\vspace{1cm}
	\includegraphics{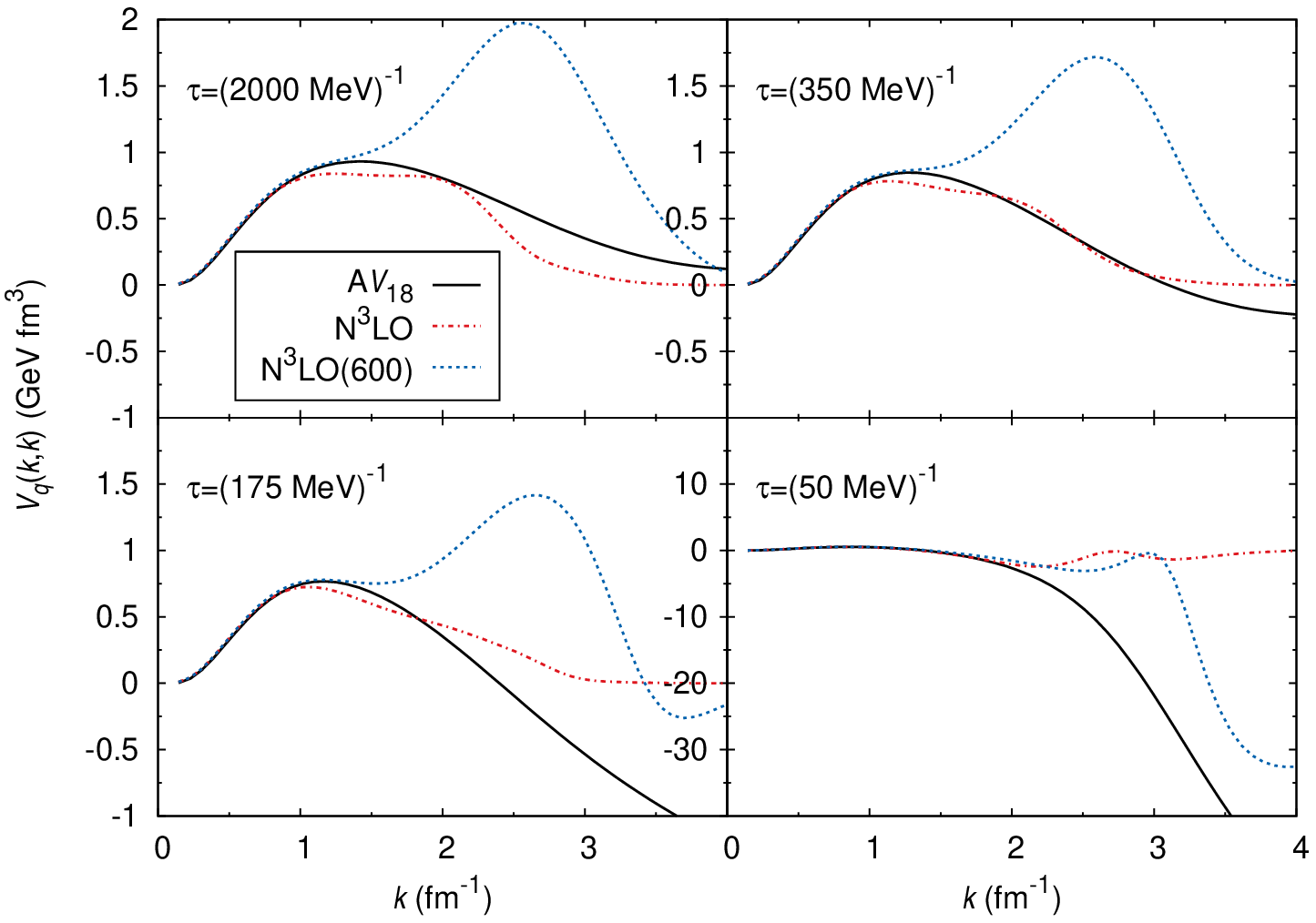}
	\vspace{1cm}
	\caption{\label{fig:3D1diag}(color online) The quantum potential in the 
	$^3\!D_1$ partial wave as the propagator is calculated for successively 
	longer imaginary times.}
\end{figure*}

\begin{figure*}[htp]
	\vspace{1cm}
	\includegraphics{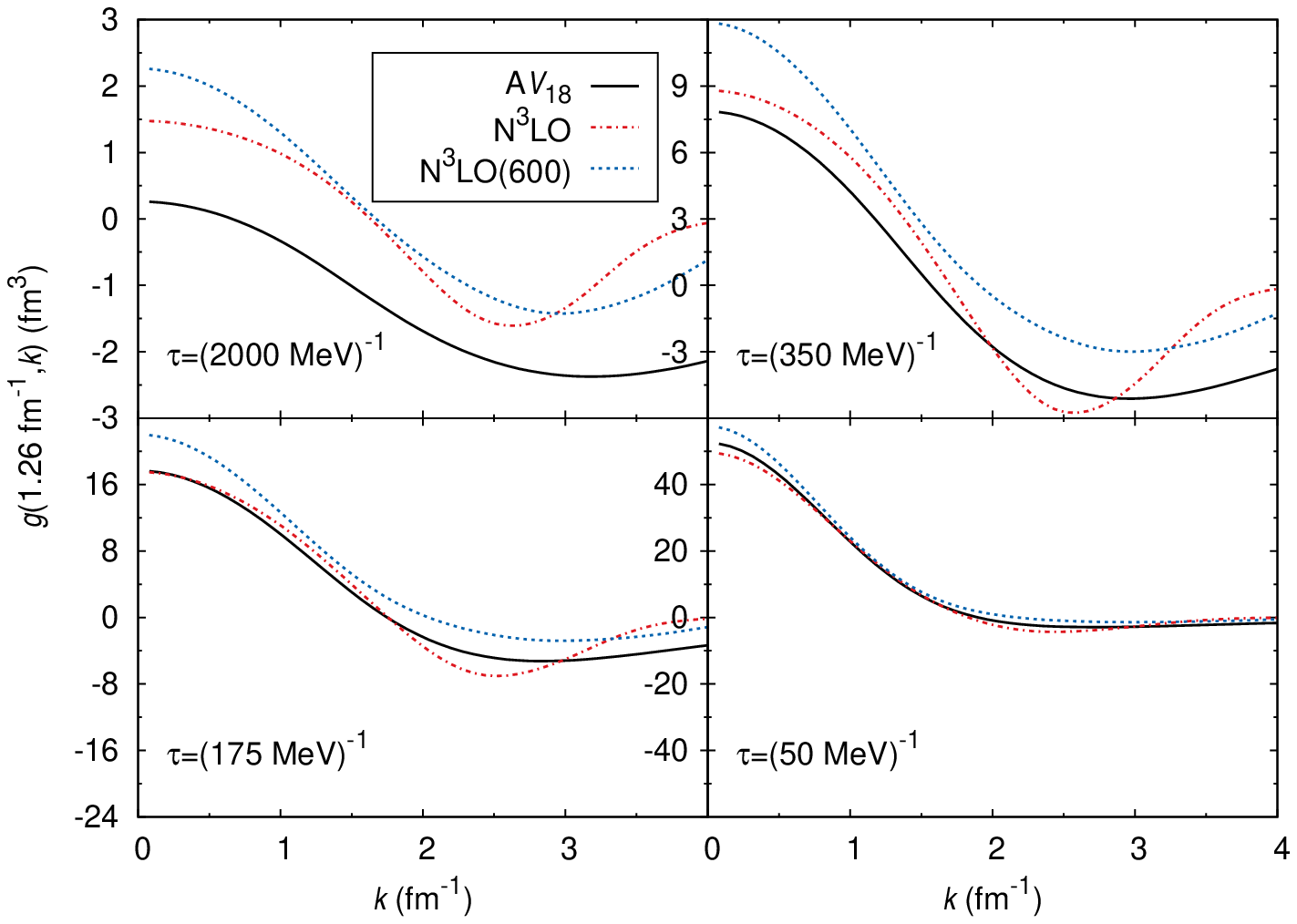}
	\vspace{1cm}
	\caption{\label{fig:1S0offdiag}(color online) The off-diagonal 
	momentum-space propagator (minus the free-particle propagator) in the 
	$^1\!S_0$ partial wave as the propagator is calculated for successively 
	longer imaginary times.}
\end{figure*}

\begin{figure*}[htp]
	\vspace{1cm}
	\includegraphics{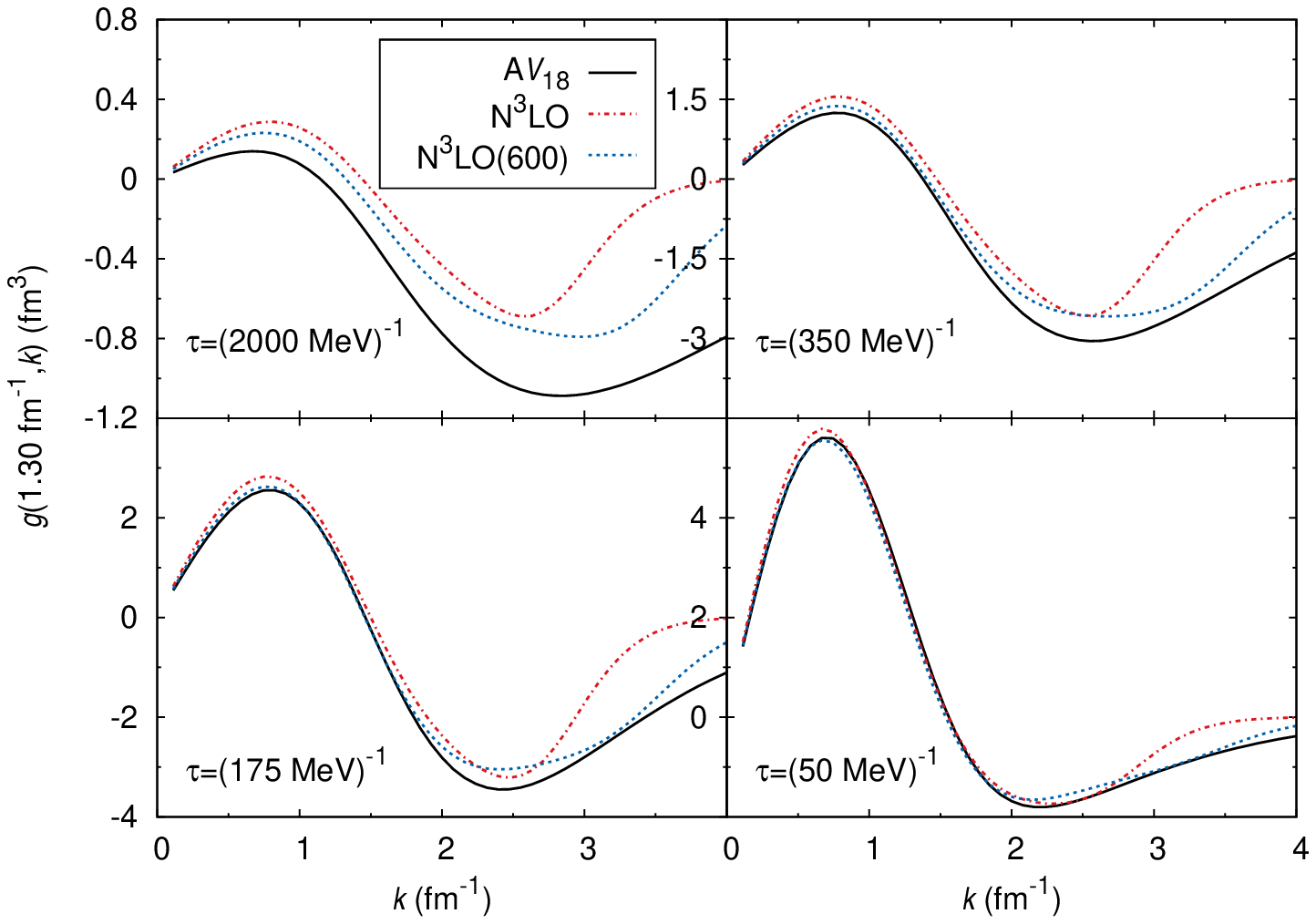}
	\vspace{1cm}
	\caption{\label{fig:3P0offdiag}(color online) The off-diagonal 
	momentum-space propagator (minus the free-particle propagator) in the 
	$^3\!P_0$ partial wave as the propagator is calculated for successively 
	longer imaginary times.}
\end{figure*}

\begin{figure*}[htp]
	\vspace{1cm}
	\includegraphics{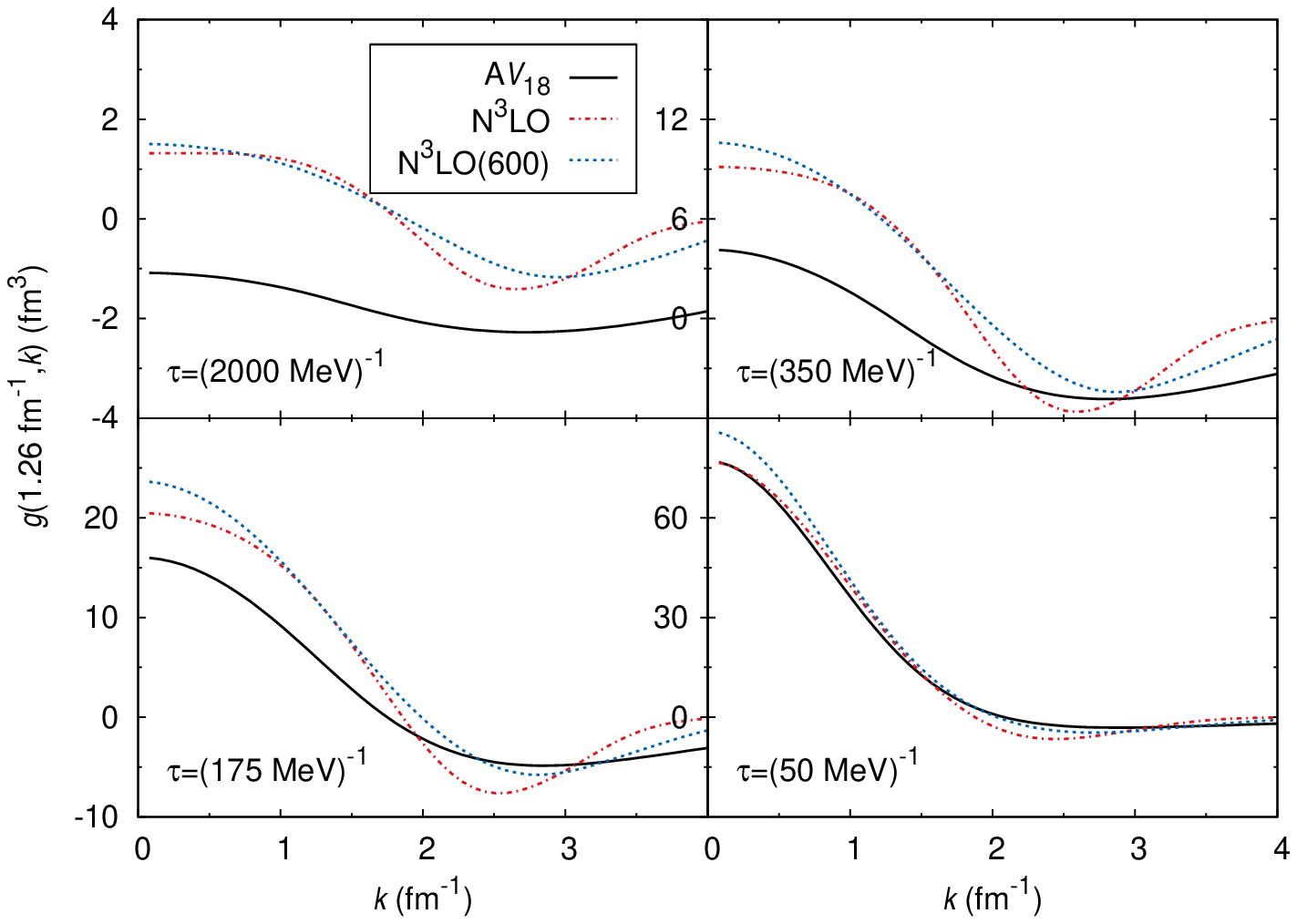}
	\vspace{1cm}
	\caption{\label{fig:3S1offdiag}(color online) The off-diagonal 
	momentum-space propagator (minus the free-particle propagator) in the 
	$^3\!S_1$ partial wave as the propagator is calculated for successively 
	longer imaginary times.}
\end{figure*}

\begin{figure*}[htp]
	\vspace{1cm}
	\includegraphics{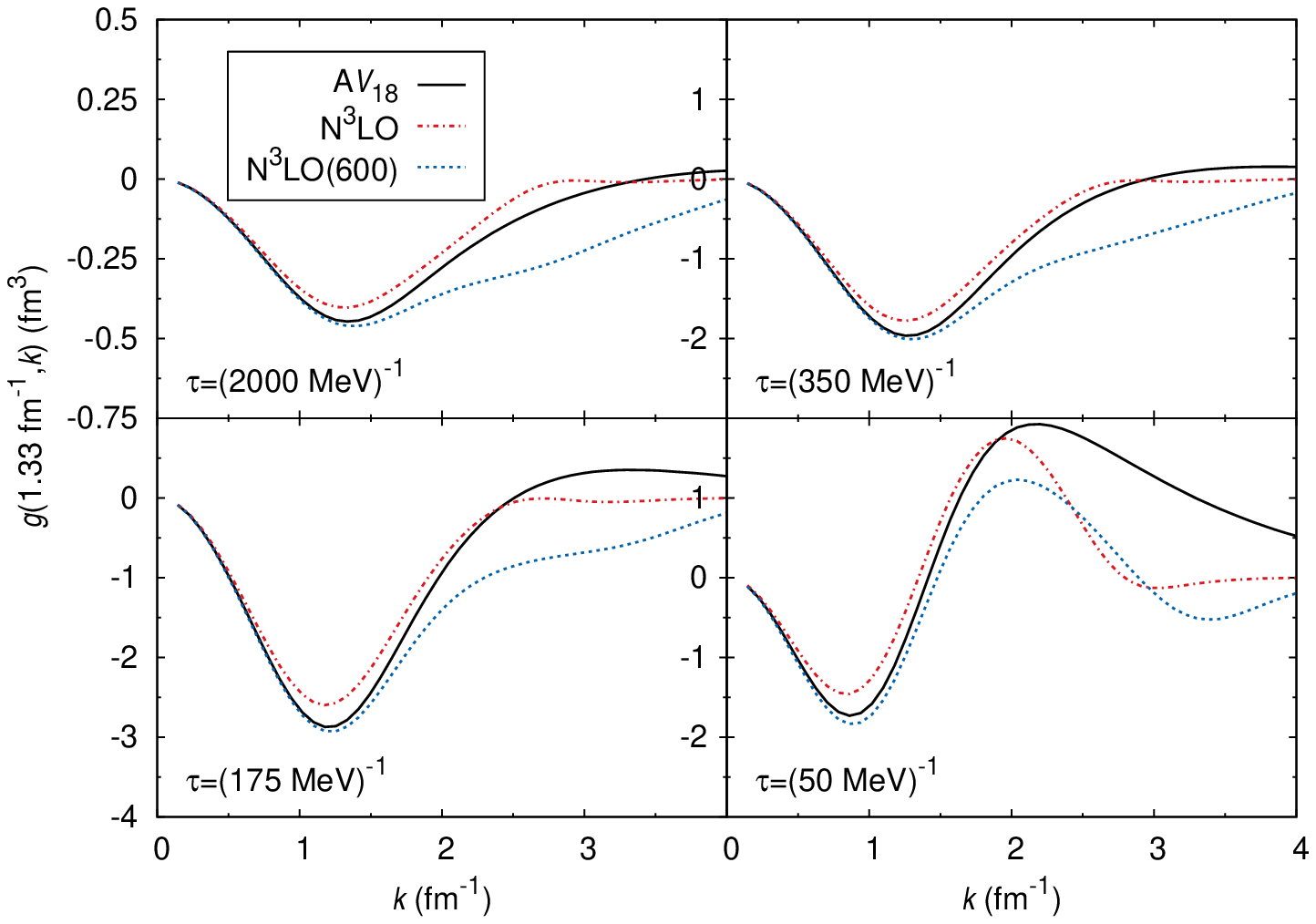}
	\vspace{1cm}
	\caption{\label{fig:3D1offdiag}(color online) The off-diagonal 
	momentum-space propagator (minus the free-particle propagator) in the 
	$^3\!D_1$ partial wave as the propagator is calculated for successively 
	longer imaginary times.}
\end{figure*}

\begin{figure*}[htp]
	\vspace{1cm}
	\includegraphics{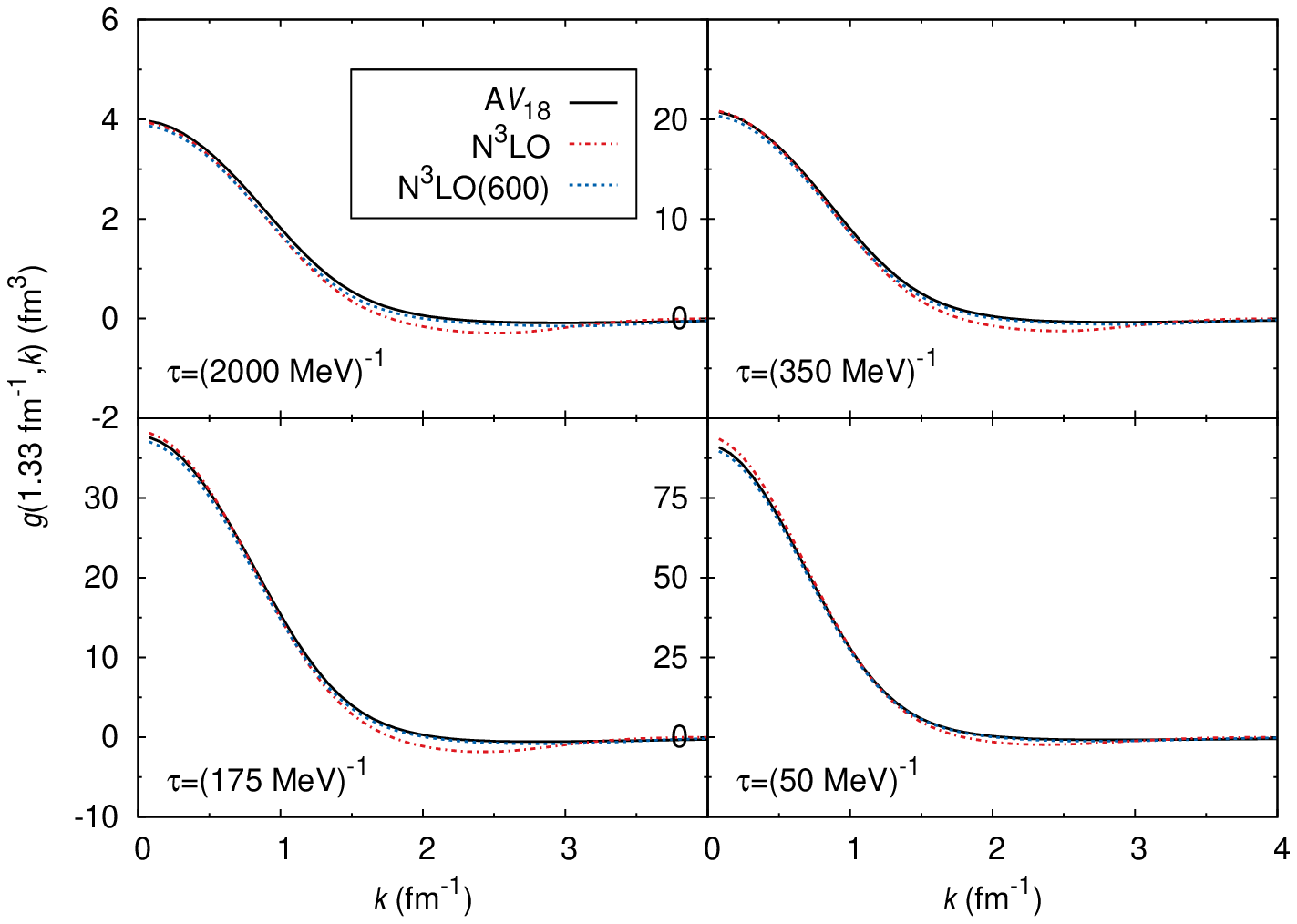}
	\vspace{1cm}
	\caption{\label{fig:3S13D1offdiag}(color online) The off-diagonal 
	momentum-space propagator (minus the free-particle propagator) in the 
	$^3\!S_1$-$^3\!D_1$ partial wave as the propagator is calculated for 
	successively longer imaginary times.}
\end{figure*}

Figures \ref{fig:1S0diag}--\ref{fig:3S13D1offdiag} demonstrate 
these findings for three potentials: Argonne $v_{18}$ 
(A$V_{18}$) \cite{Wiringa:1994wb}, 
N$^3$LO, and N$^3$LO(600) \cite{Entem:2003ft}.  For the diagonal cases, 
where we take $k=k^\pr$, we define
a quantum potential, 
$V_{q}(k,k;\tau)$ through the equation
\begin{equation}
\matrixel{k}{e^{-H\tau}}{k}
=\matrixel{k}{
e^{-H_0\frac{\tau}{2}}
e^{-V_q(k,k;\tau)\tau}
e^{-H_0\frac{\tau}{2}}}{k},
\end{equation}
with $H_0$ the free-particle Hamiltonian 
\begin{equation}
H_0=T=\frac{p^2}{2m}.
\end{equation}
In the off-diagonal cases ($k\ne k^\pr$) we choose a particular $k$ value
and plot against $k^\pr$.  For visual comparison, we subtract the 
free-particle propagator, $g(k,k^\pr)-g_0(k,k^\pr)$, since at the point 
$k=k^\pr$, the kinetic energy component is large and obscures
the result.

Figures \ref{fig:1S0diag}--\ref{fig:3D1diag}
show the quantum potential in the singlet ($^1\!S_0$),
uncoupled triplet ($^3\!P_0$), and coupled triplet ($^3\!S_1$ and 
$^3\!D_1$) channels.  The imaginary times chosen correspond to a typical 
time step used in Green's function Monte Carlo calculations, 
$\tau=(2000\text{ MeV})^{-1}$ and imaginary times that roughly correspond 
to center-of-mass energies of 350 MeV, 175 MeV, and 50 MeV.  As we have 
discussed above, the imaginary time of $\tau=(50\text{ MeV})^{-1}$ is the 
time at which the effects attributable to energies of 175 MeV and above are
effectively integrated out.  It is interesting to note that the agreement of
the quantum potentials is only good up to $k\approx2\text{ fm}^{-1}$: this 
is the approximate momentum value, $k$, one would associate with the
corresponding kinetic energy: $\frac{\hbar^2k^2}{2m}=175\text{ MeV}$.
This relationship (better and better agreement --- but only up to some
cut off --- as the potential is evolved) is precisely what is found in 
similarity renormalization group and $V_{\text{low }k}$ approaches.  

It is tempting to interpret $\tau$ as an evolution parameter for the quantum
potential in the same sense that the similarity renormalization group 
approach has $s$ or $\lambda$ (see, for example, \cite{bogner:2006pc}).
However, it is not clear if a direct comparison is at all trivial.  In the 
similarity renormalization group approach, the evolution parameter $s$ (and
 therefore $\lambda$, since $\lambda=1/s^{1/4}$) is defined through the 
rather simple evolution equation for the potential
\begin{equation}
\label{eq:srgevolution}
\frac{dH_s}{ds}=\frac{dV_s}{ds}=[[G_s,H_s],H_s],
\end{equation}
where $G_s$ is a Hermitian operator that generates the transformation.  
($G_s$ is often chosen to be $T$, the kinetic energy).  If we view our
quantum potential akin to the similarity renormalization group's $V_s$,
our defining equation is
\begin{equation}
e^{-H\tau}=
e^{-T\frac{\tau}{2}}
e^{-V_q(\tau)\tau}
e^{-T\frac{\tau}{2}} \,.
\end{equation}
We can expand this using the
Baker-Campbell-Hausdorff relation which gives an infinite series
of nested commutators. The lowest order terms in an expansion in $\tau$ are
\begin{equation}
V_q(\tau) = V -\tfrac{\tau^2}{12} \left (
[V,[V,T]]-\tfrac{1}{2}[T,[T,V]]\right ) + \cdots
\end{equation}
where the double commutators are suggestive of Eq. (\ref{eq:srgevolution}),
but clearly not the same. The differential equation satisfied by $V_q(\tau)$
is not a simple, compact expression.

Figures \ref{fig:1S0offdiag}-\ref{fig:3S13D1offdiag}, show the off-diagonal 
elements of the singlet ($^1\!S_0$), uncoupled triplet, ($^3\!P_0$), and 
coupled triplet ($^3\!S_1$, $^3\!D_1$, and $^3\!S_1$-$^3\!D_1$) channel 
propagators.  As discussed above, they are shifted by the free-particle 
result to make comparisons easier.  What we can see from these figures is a
general trend towards universality with at least two caveats.  
First, the propagators tend to converge most rapidly around the point where
$k=k^\pr$.  Our interpretation of this result is that low momentum transfer
behavior is constrained by the phase-shift equivalence of the various 
potentials whereas higher momentum transfer behavior is not.  Second, some
channels converge better than others.  For example, Fig. 
\ref{fig:3D1offdiag} has still not converged at $\tau=(50\text{ MeV})^{-1}$,
and indeed, does not appear to converge well until 
$\tau=(10\text{ MeV})^{-1}$.  This may be due to this channel's sensitivity
to the tensor part of the interaction which the different potentials
treat differently.  These differences may point to true, quantifiable 
distinctions between the potentials.

Even though the potentials give propagators with the same sort of long
imaginary time behavior, the many-body physics of the nucleus may not
allow the use of the propagators in this regime. As mentioned
above, in Green's
function Monte Carlo calculations, the imaginary time step needed to
accurately approximate the many-body Green's function by the pair-product
is $\tau=(2000\text{ MeV})^{-1}$. For larger time steps, commutator terms
in the Trotter breakup spoil the approximation. Since, in the pair product
approximation, these commutators occur only when three nucleons are
close together, this indicates that, three-body effects will also
be important. That is, to be able to use the propagators in the
limit where they become model independent, would likely require that
three- and more-body terms in both the interaction and the propagators
be included. This likely means that use of potentials like
$V_{\text{low }k}$ for many-nucleon calculations will need to include
many-body interactions.

\section{Quantum Monte Carlo with non-local potentials}
We now turn to the central parts of the propagators for the non-local 
N$^3$LO and N$^3$LO(600) potentials we have been considering throughout.
Since
the central part of the propagator is sampled in a quantum Monte
Carlo calculation it should be positive-definite to avoid sign problems.

\begin{figure}[htp]
	\includegraphics{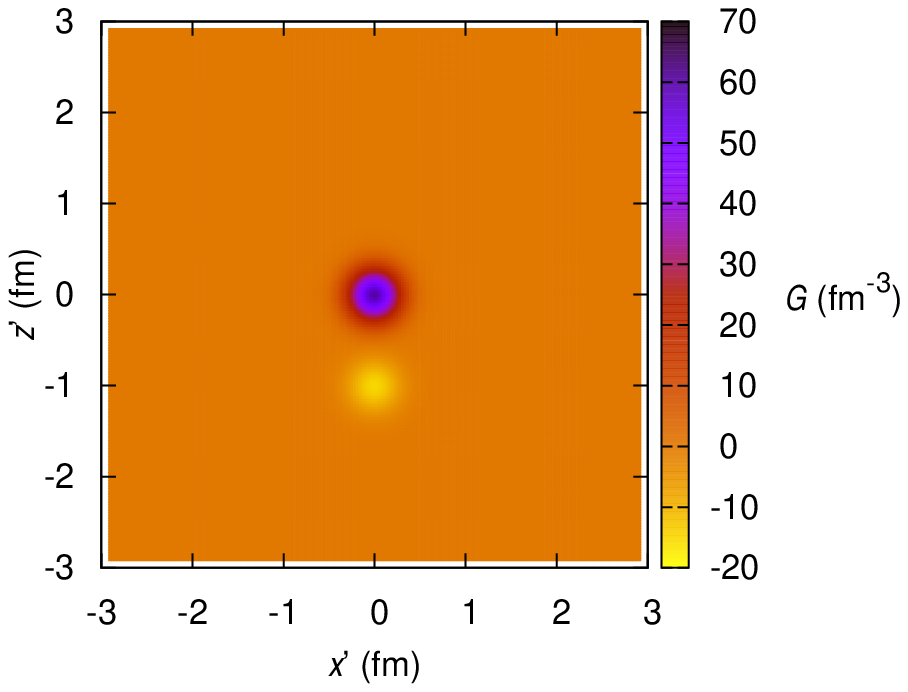}
	\caption{\label{fig:n3locentral2000}(color online) The N$^3$LO central
	propagator for initial separation, $r=0.5$ fm, and imaginary time,
	$\tau=(2000\text{ MeV})^{-1}$.}
\end{figure}

\begin{figure}[htp]
	\includegraphics{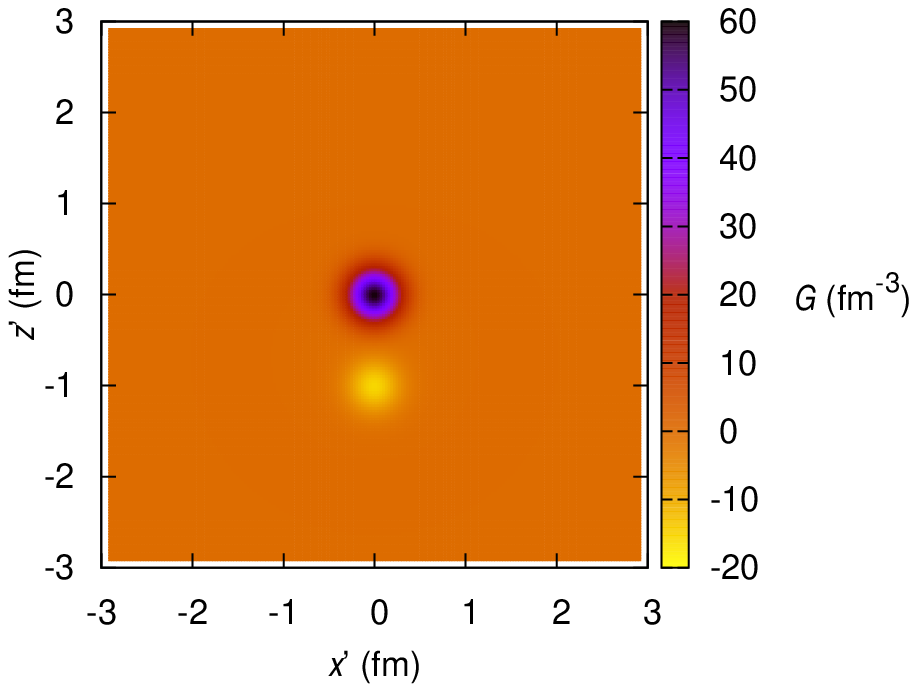}
	\caption{\label{fig:n3lo600central2000}(color online) The N$^3$LO(600)
	central propagator for initial separation, $r=0.5$ fm, and imaginary 
	time, $\tau=(2000\text{ MeV})^{-1}$.}
\end{figure}

\begin{figure}[htp]
	\includegraphics{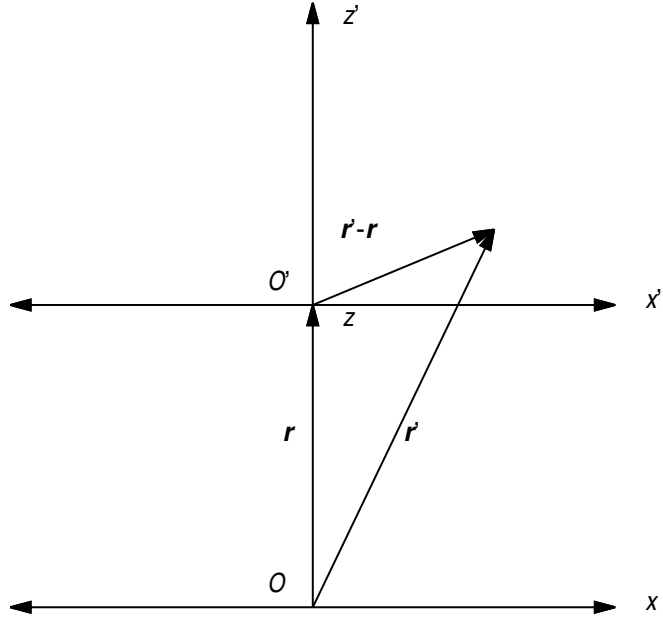}
	\caption{\label{fig:coordinatesystem}The coordinate system used to 
	visualize the central propagators.  $O$ is the original origin, and 
	$O^\pr$ the shifted origin such that $r=r^\pr$ corresponds to the 
	shifted origin.}
\end{figure}

Figures \ref{fig:n3locentral2000} and \ref{fig:n3lo600central2000} plot the
central part of the propagator in real space, where the coordinates 
$x^\prime$ and
$z^\prime$ are such that the origin corresponds to the final separation 
equal to
the initial separation.  That is, the relative coordinates are equal: 
$r=r^\pr$.  The precise transformation between the original coordinates,
$r$, $r^\pr$, and $\theta^\prime$ and the new coordinates, $x^\prime$ and 
$z^\prime$ is 
given by
\begin{subequations}
\begin{eqnarray}
r^{\pr\,2}&=&x^{\prime\,2}+(r+z^\prime)^2\\
\cos\theta^\prime&=&\frac{r+z^\prime}{\sqrt{x^{\prime\,2}+(r+z^\prime)^2}},
\end{eqnarray}
\end{subequations}
and can be visualized in Fig. \ref{fig:coordinatesystem}.

The structure we can see is a Gaussian-like peak about the initial 
separation as well as an antisymmetric trough at the position that 
corresponds to the two nucleons undergoing a position interchange: 
$r^\pr=-r$.
The
antisymmetric point is built in from tracing over the spins and isospins as
in Eq.~(\ref{eq:centralprop}), and would be present even for Argonne 
$v_{18}$.  
This point gives no extra difficulty --- since it comes
from the fermion character of the nucleons, it will be dealt with
in the same way that the fermion sign problem is dealt with in
Green's function or auxiliary field diffusion Monte Carlo. That is,
a path constraint \cite{wiringa2000} is imposed that eliminates the fermion
sign problem. The constraint can then be released and forward walking
steps taken \cite{wiringa2000,pieper2002} to improve the results and
check the effect of the constraint.

\begin{figure}[htp]
	\includegraphics{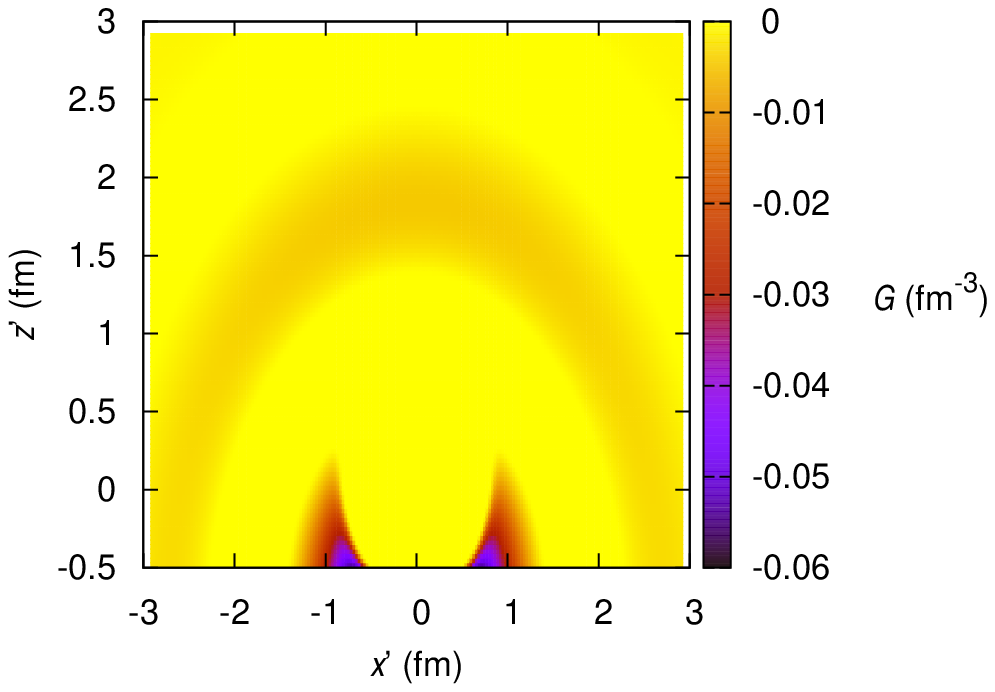}
	\caption{\label{fig:n3lonegcentral2000}(color online) The negative parts
	of the N$^3$LO central propagator for initial separation, $r=0.5$ fm, 
	and imaginary time, $\tau=(2000\text{ MeV})^{-1}$.}
\end{figure}

\begin{figure}[htp]
	\includegraphics{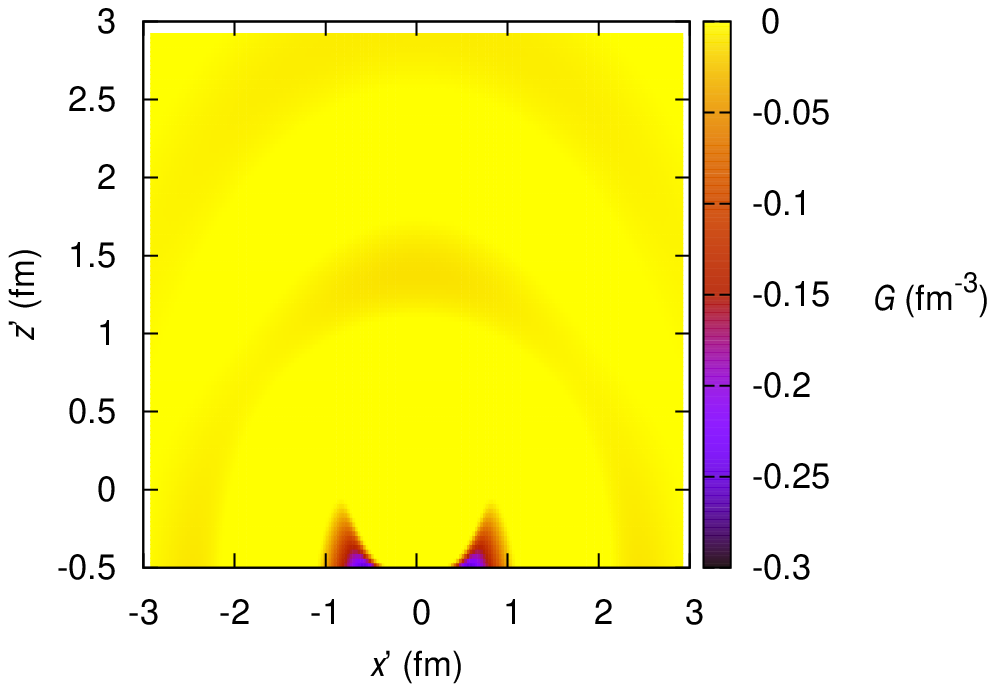}
	\caption{\label{fig:n3lo600negcentral2000}(color online) The negative 
	parts of the N$^3$LO(600) central propagator for initial separation, 
	$r=0.5$ fm, and imaginary time, $\tau=(2000\text{ MeV})^{-1}$.}
\end{figure}

\begin{figure}[htp]
	\vspace{1cm}
	\includegraphics{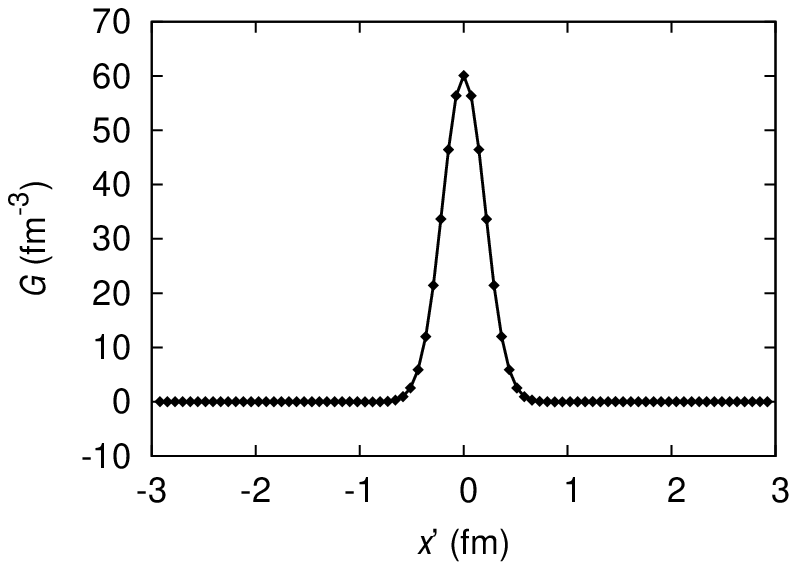}
	\vspace{1cm}
	\caption{\label{fig:n3locentralslice2000}A typical
	slice through the N$^3$LO central propagator for initial separation,
	$r=0.5$ fm, and imaginary time, $\tau=(2000\text{ MeV})^{-1}$.}
\end{figure}

\begin{figure}[htp]
	\vspace{1cm}
	\includegraphics{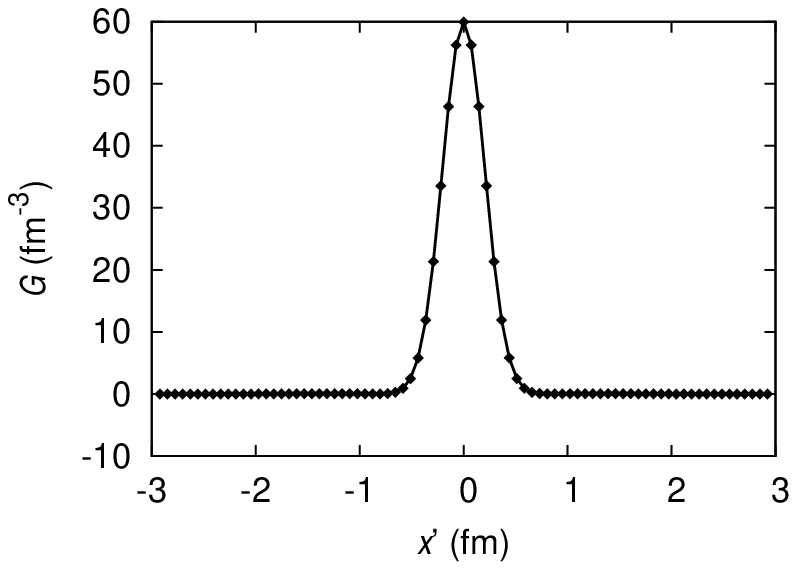}
	\vspace{1cm}
	\caption{\label{fig:n3lo600centralslice2000}A typical
	slice through the N$^3$LO(600) central propagator for initial 
	separation, $r=0.5$ fm, and imaginary time, 
	$\tau=(2000\text{ MeV})^{-1}$.}
\end{figure}

However, if we zoom in on the shifted origin, as in Figs. 
\ref{fig:n3lonegcentral2000}, and \ref{fig:n3lo600negcentral2000}, setting 
any positive parts of the propagator to zero, and make sure we are clear of
the antisymmetric region, we find that the propagators appear to be 
``ringing'', much like Friedel oscillations.
These negative parts may make it more difficult to perform quantum Monte Carlo
calculations and keep the sign problem under control.  However,
these negative parts are quite small, of order $10^{-1}\text{ fm}^{-3}$,
whereas the peak of the propagator is of order $10^2\text{ fm}^{-3}$.  In
fact, a typical slice through the propagator in the $x^\prime$ direction 
looks
like  Figs. \ref{fig:n3locentralslice2000} and 
\ref{fig:n3lo600centralslice2000}.  In many cases, the negative parts are
negligible.

A straightforward method to take the small negative regions
into account, is to first set any negative part (not associated with 
the antisymmetry) to zero, run a quantum Monte Carlo calculation until it 
converges, and then add the negative parts back in, in a perturbative 
fashion, using forward walking exactly as for the fermion sign
problem.
The extra sign changes from the propagator are handled in the same
way as sign changes from the fermion character.

We can estimate the fraction of walkers in the initial time
step that may be given negative weights by comparing the integral of the
absolute value of the propagators to the integral of the propagators.  That 
is, we estimate the fraction of walkers with negative weights, $f$ by
\begin{equation}
f=\frac{\int d^3r^\prime\tfrac{1}{2}\left[|G(\bm{r},\bm{r}^\prime;\tau)|-
G(\bm{r},\bm{r}^\prime;\tau)\right]}
{\int d^3r^\prime|G(\bm{r},\bm{r}^\prime;\tau)|}.
\end{equation} 
We calculate
the integral over the upper half volume to exclude the interchange.
For N$^3$LO with an initial separation of 1.0 fm, we find 
$f\sim\mathcal{O}(10^{-2})$.  If we now take the 
very conservative estimate for the alpha particle
that all six pairs may be this close at one time and
that the negative weights are acceptable so 
long as the fraction of walkers with negative weights is less than $1/e$, 
we find we can take approximately ten steps for N$^3$LO [with time step
$(2000\text{ MeV})^{-1}$].
Green's function Monte Carlo typically uses forward walking of about 10 to 
20 steps of $(2000\text{ MeV})^{-1} $\cite{pieper2002,wiringa2000}. Therefore
forward walking will allow us to remove any bias from the negative
parts of the propagator. This can be compared to forward walking keeping
the propagator constraint, but releasing the fermion constraint to separate
the two effects.

In the above analysis, we have assumed that the imaginary
time step used will be the same as that used in Green's function
Monte Carlo calculations with the Argonne family of potentials. Since
the N$^3$LO potentials are softer, the relevant commutators terms will
be smaller, and longer time steps may be possible. For longer time
steps there is much less ringing, and the calculations will be substantially
easier.

\section{Conclusions}

We have shown how to calculate the imaginary time
pair propagators needed for quantum
Monte Carlo calculations of nuclei and nuclear matter using non-local
potentials in momentum space.
The method is general
enough to handle any non-local potential in momentum or real space, but in 
this paper, we focus on those derived from effective field theory (N$^3$LO).

We find that the propagators display a universal behavior at large imaginary
times, consistent with our expectations from renormalization group methods
and the fact
that the potentials are  phase-shift equivalent, meaning that they
reproduce the scattering data at or below laboratory energies of 350 MeV.  

The central propagators sampled during Monte Carlo simulations for local
potentials are expected
to be positive-definite.  Without this property, sign problems can develop.
We find that for N$^3$LO with a 500 MeV or 600 MeV cutoff in momentum space,
the central propagator is not positive definite.  However, the negative
parts consist of ``rings'' reminiscent of Friedel oscillations and their
magnitude is quite small compared with the overall shape of the central
propagator. Since these potentials were developed in momentum space,
no attempt was made to influence their behavior in position space.
It may be possible to modify the N$^3$LO potentials in such a 
way that they continue to reproduce the Nijmegen data with a low $\chi^2$,
are still relatively soft,
but have reduced ringing behavior.  A modification
of the
choice of the regulator function used in the calculation of the N$^3$LO 
potentials:
$V(k,k^\pr)\rightarrow V(k,k^\pr)e^{-(k/\Lambda)^{2\nu}}
e^{-(k^\pr/\Lambda)^{2\nu}},$ where $\Lambda$ is the cutoff value, and $\nu$
is the order of the calculation, ($\nu=4$ for N$^3$LO) may help.
In any case, quantum 
Monte Carlo calculations should still be possible by using
a modified path constraint as described above, and we are implementing these
calculations.

While we have concentrated on calculating the imaginary time pair propagators
from phenomenological potentials, it is amusing to note that since
the few-body
imaginary time propagators are simply imaginary-time correlations of the
appropriate nucleon operators, they
might eventually be directly extracted from lattice QCD calculations.

\begin{acknowledgments}
The authors would like to thank J. Carlson, R. J. Furnstahl, and S. Gandolfi
for helpful conversations.
This work was supported by the National
Science Foundation under Grant No. PHY-1067777.
\end{acknowledgments}

%

\end{document}